%% file: ms.tex
\documentclass[10pt,conference]{IEEEtran}
\IEEEoverridecommandlockouts
\usepackage{cite}
\usepackage[utf8]{inputenc}
\usepackage{amsmath,amssymb,amsfonts,amsthm}
\usepackage{algorithmic}
\usepackage{graphicx}
\usepackage{textcomp}

\usepackage[capitalize]{cleveref}
\usepackage{lazyeqn}
\usepackage{shortsym}
\usepackage{siunitx}
\usepackage{pgfplots}
\pgfplotsset{compat=newest}
\usepackage{setspace}

\DeclareMathOperator*\argmin{arg \, min}		
\DeclareMathOperator*{\minimize}{minimize}
\DeclareMathOperator{\prox}{prox}

\DeclareMathOperator{\refold}{refold}
\DeclareMathOperator{\shrink}{shrink}
\DeclareMathOperator{\mean}{mean}

\DeclareMathOperator{\Tv}{Tv}

\newtheorem{remark}{Remark}

\IEEEsettopmargin{t}{.8in}
\IEEEsettextwidth{0.75in}{0.75in}

\def\BibTeX{{\rm B\kern-.05em{\sc i\kern-.025em b}\kern-.08em
    T\kern-.1667em\lower.7ex\hbox{E}\kern-.125emX}}
\begin{document}

\setstretch{0.938}

\title{Tensor Completion for Radio Map Reconstruction using Low Rank and Smoothness
\thanks{This work is supported by the Federal Ministry of Education and Research of the Federal Republic of Germany (BMBF) in the framework of the project 5G NetMobil with funding number 16KIS0691. The authors alone are responsible for the content of the paper.}}

\author{\IEEEauthorblockN{Daniel Schäufele\IEEEauthorrefmark{1}, Renato L. G. Cavalcante\IEEEauthorrefmark{1}\IEEEauthorrefmark{2},
Slawomir Stanczak\IEEEauthorrefmark{1}\IEEEauthorrefmark{2}
}
\IEEEauthorblockA{\IEEEauthorrefmark{1}\textit{Fraunhofer Heinrich Hertz Institute}, Berlin, Germany \\
\IEEEauthorrefmark{2}\textit{Technische Universität Berlin}, Berlin, Germany \\
\{daniel.schaeufele, renato.cavalcante, slawomir.stanczak\}@hhi.fraunhofer.de \\
}
}

\newcommand\copyrighttext{%
  \footnotesize \textcopyright 2012 IEEE. Personal use of this material is permitted.
  Permission from IEEE must be obtained for all other uses, in any current or future
  media, including reprinting/republishing this material for advertising or promotional
  purposes, creating new collective works, for resale or redistribution to servers or
  lists, or reuse of any copyrighted component of this work in other works.
}
\newcommand\copyrightnotice{%
\begin{tikzpicture}[remember picture,overlay]
\node[anchor=south,yshift=10pt] at (current page.south) {\fbox{\parbox{\dimexpr\textwidth-\fboxsep-\fboxrule\relax}{\copyrighttext}}};
\end{tikzpicture}%
}

\maketitle
\copyrightnotice

\begin{abstract}
    Radio maps are important enablers for many applications in wireless networks, ranging from network planning and optimization to fingerprint based localization.
    Sampling the complete map is prohibitively expensive in practice, so methods for reconstructing the complete map from a subset of measurements are increasingly gaining attention in the literature.
    In this paper, we propose two algorithms for this purpose, which build on existing approaches that aim at minimizing the tensor rank while additionally enforcing smoothness of the radio map.
    Experimental results with synthetic measurements derived via ray tracing show that our algorithms outperform state of the art techniques.
\end{abstract}


\section{Introduction}
\label{sec:introduction}

Accurate and reliable characterization of radio frequency environments is a key ingredient in the design and operation of modern radio networks.
Knowledge of the spatial distribution of the path loss is essential for many applications in wireless networks, particularly those expected in future 5G systems.
One common application in which accurate estimation of path loss maps is required is proactive resource allocation, where the knowledge of future propagation conditions along a user's trajectory is utilized to allocate resources\cite{bui2016anticipatory}.
Path loss maps can also be used for fingerprint-based localization schemes, where the geographical locations of users are estimated based on their measured channel states \cite{hu_efficient_2013,nikitaki_efficient_2012}.

For the above reasons, path loss estimation has gained a great deal of attention from both the academia and the industry, but accurate path loss estimation in wireless networks remains a challenging task to date \cite{phillips_survey_2013}. A plethora of theoretical and empirical path loss models have been proposed by researchers over the years, and, although some have been proved to be useful in selected applications, many existing models fail to deliver good predictive performance in real world settings \cite{phillips_efficacy_2011}. As a result, data-driven methods based on learning tools are increasingly gaining attention, and they have been shown to produce results with higher reliability and accuracy than purely model-driven methods\cite{kasparick_kernel-based_2016}.

Many of these data-driven methods for radio map reconstruction utilize matrix completion algorithms\cite{hu_efficient_2013,nikitaki_efficient_2012,chouvardas_method_2016,claude_efficient_2017}, and, in the field of image and video processing, extensions of these matrix reconstruction methods to higher-order tensors have been shown to provide superior performance\cite{liu2013tensor,wang2014low}.
However, tensor methods have rarely been applied to radio map reconstruction problems.

Against this background, we propose novel approaches for radio map reconstruction based on tensor reconstruction algorithms.
One of the main advantages of the proposed algorithms is that they can easily take into account information such as path loss in multiple heights, frequency bands, and towards multiple base stations, among others.
In addition, the proposed algorithms exploit the fact that radio maps are typically smooth to increase the quality of the reconstruction.
We evaluate the performance of the proposed algorithms by using synthetic measurements that were obtained from ray tracing simulations performed during the METIS2020 project\cite{metis_d61}.

The remainder of this paper is organized as follows.
In \cref{sec:preliminaries} we introduce standard results in mathematics that are crucial to the algorithms we propose.
In \cref{sec:algorithm} we derive novel algorithms to solve the aforementioned problems.
Numerical experiments are given in \cref{sec:experiments}.

{\bf Notation:} Scalars, column vectors, matrices, and tensors are denoted by italic lower case letters $x$, bold lower case letters $\Vx$, italic upper case letters $X$, and bold upper case letters $\VX$, respectively. The element $(i_1, i_2, \dots, i_N)$ of a tensor $\VX \in \IR^{n_1 \times \dots \times n_N}$ is denoted by $x_{i_1 i_2 \dots i_N}$.
The vector space containing $N$th order tensors is denoted by $\CT := \IR^{n_1 \times \dots \times n_N}$.
By equipping this vector space with the inner product $\abr{\VX, \VY} := \sum_{i_1=1}^{n_1} \sum_{i_2=1}^{n_2} \dots \sum_{i_N=1}^{n_N} x_{i_1 i_2 \dots i_N} y_{i_1 i_2 \dots i_N}$ and its induced norm $\Norm{\VX} := \sqrt{\abr{\VX,\VX}}$ we obtain a Hilbert space of tensors.
The high-order generalizations of columns and rows of a matrix are called fibers, which are extracted by fixing all coordinates except for one.
In particular, mode-$m$ fibers are denoted by $\Vx_{i_1\dots i_{m-1}:i_{m+1}\dots i_N}$, where a colon in the index is used to indicate all elements of a mode.
To compute the mode-$m$ unfolding of a tensor, we rearrange all mode-$m$ fibers of the tensor as the columns of a matrix. More formally, $X_{(m)} \in \IR^{n_m \times I_m}$ denotes the mode-$m$ unfolding of the tensor $\VX$, where $I_m = \prod_{\substack{k=1\\k\ne m}}^{N} n_k$ and the tensor element $(i_1, \dots, i_N)$ is mapped to the matrix element $(i_m, j)$, with $j = 1+ \sum_{\substack{k=1\\k \ne m}}^{N} (i_k - 1) J_k$ and $J_k = \prod_{\substack{\ell=1\\\ell \ne m}}^{k-1} n_\ell$.
The $m$-mode multiplication of a tensor $\VX \in \IR^{n_1 \times \dots \times n_N}$ and a matrix $Y \in \IR^{J \times n_m}$ is denoted by $\VX \times_m Y,$ and it is defined by the following relation based on the unfoldings: $\VZ = \VX \times_m Y \ \Leftrightarrow\ Z_{(m)} = Y X_{(m)}$.
The nuclear norm of a matrix $X$ is defined as $\Norm{X}_* := \sum_i \sigma_i(X)$, where $\sigma_i(X)$ denotes the $i$th largest singular value of $X$.
The class of all lower semi-continuous convex functions from an arbitrary real Hilbert space $\CH$ to $(-\infty, \infty]$ that are not identically equal to $+\infty$ are denoted by $\Gamma_0(\CH)$.

\section{Preliminaries}
\label{sec:preliminaries}

The proposed algorithms are heavily based on the Douglas-Rachford splitting method, so we briefly review this iterative method in \cref{sec:dr}. Then, in \cref{sec:rank}, we present the algorithm introduced in \cite{gandy_tensor_2011}, which is the algorithm we improve in \cref{sec:algorithm} by considering specific features of radio maps.

\subsection{Douglas-Rachford splitting}
\label{sec:dr}

Given an arbitrary Hilbert space $\CH$, the objective of the Douglas-Rachford splitting algorithm is to minimize the sum of two functions $f,g \in \Gamma_0(\CH)$ satisfying some additional very mild assumptions \cite[Corollary~28.3]{bauschke2017convex} that are valid in the applications described in the next sections.
This algorithm generates a sequence $(x_n)_{n \in \IN}$ given by
\begin{align}
    x_{n+1} := x_n + t_n \rbr{\prox_{\gamma f} \rbr{2 \prox_{\gamma g} x_n - x_n} - \prox_{\gamma g} x_n},
    \label{eq:dr_splitting}
\end{align}
where $t_n \in [0,2]$ is the step size and
\begin{align}
    \prox_{\gamma f} : \CH \to \CH: x \mapsto \argmin_{y \in \CH} \rbr{f(y) + \frac{1}{2 \gamma} \Norm{x - y}_{\CH}^{2} }
    \label{eq:prox}
\end{align}
denotes the proximal map of a function $f \in \Gamma_0(\CH)$ with regularization parameter $\gamma \in (0, \infty)$\cite[p.~211]{bauschke2017convex}.
For convergence properties of the recursion in \eqref{eq:dr_splitting}, we refer the readers to \cite[Section~28.3]{bauschke2017convex}.

Although the Douglas-Rachford method as shown above is used to minimize the sum of two functions in $\Gamma_0(\CH)$, we can straightforwardly modify the algorithm to incorporate more than two functions\cite{gandy_tensor_2011}.
Consider for concreteness the following optimization problem in the tensor space $\CT$:
\begin{align}
    \minimize_{\VX \in \CT} \sum_{i=1}^{M} f_i(\VX)
    \label{eq:min_sum_mult}
\end{align}
where $f_i \in \Gamma_0(\CT)$ for each $i \in \cbr{1, \dots, M}$.
By defining the Hilbert space $\CH_0 := \CT \times \CT \times \dots \times \CT$ ($M$-fold Cartesian product) with elements $\VX = (\VX_1, \dots, \VX_M)$ and inner product $\abr{\VX, \VY}_{\CH_0} := \frac{1}{M} \sum_{i=1}^{M} \abr{\VX_i,\VY_i}$, we can obtain a minimizer of \eqref{eq:min_sum_mult} by solving
\begin{align}
    \minimize_{\VZ \in \CH_0} f(\VZ) + g(\VZ),
    \label{eq:minimize_dr_product}
\end{align}
where $f$ and $g$ are given by
\begin{align}
    f &: \CH_0 \to (-\infty, \infty] : \VZ \mapsto \sum_{i=1}^{M} f_i(\VZ_i) \label{eq:dr_product_f}\\
    g &: \CH_0 \to (-\infty, \infty] : \VZ \mapsto \iota_D(\VZ), \label{eq:dr_product_g}
\end{align}
where
\begin{align}
    \iota_D(\VZ) = \begin{cases}0, &\text{if } \VZ_1 = \dots = \VZ_M \\ +\infty &\text{otherwise}.\end{cases}
\end{align}

We assume that the conditions for the convergence of the Douglas-Rachford splitting method are satisfied and refer the readers to \cite[Corollary~28.3]{bauschke2017convex} for more details.
Note that $\VX^{\star}$ is a solution to \eqref{eq:min_sum_mult} if and only if $(\VZ_1^{\star}, \dots, \VZ_M^{\star})$ is a solution to \eqref{eq:minimize_dr_product} and $\VX^{\star} = \VZ_1^{\star} = \dots = \VZ_M^{\star}$.

To solve \eqref{eq:minimize_dr_product} using the iteration in \eqref{eq:dr_splitting}, we can use the proximal maps of $f$ and $g$ given by \cite{gandy_tensor_2011}
\begin{align}
    \prox_{\gamma f} \VZ &= \rbr{\prox_{\gamma f_1} \VZ_1, \dots, \prox_{\gamma f_M} \VZ_M} \label{eq:prox_f}\\
    \prox_{\gamma g} \VZ &= \rbr{\mean(\VZ), \dots, \mean(\VZ)}, \label{eq:prox_g}
\end{align}
where $\mean(\VZ) := \frac{1}{M} \sum_{i=1}^{M} \VZ_i$.

\subsection{Convex approximation of the rank minimization problem}
\label{sec:rank}

Suppose we have the problem of completing an $N$th-order tensor $\VX \in \CT$ of low rank from a subset of $p$ given entries, which we express by $\CA(\VX) = \Vb$, where $\CA: \CT \to \IR^{p}$ is a linear map that maps the values at positions of the tensor for which samples are available (indicated by the set $\Omega \subset \cbr{1, \dots, n_1} \times \dots \times \cbr{1, \dots, n_N}$) into a vector and $\Vb \in \IR^p$ is the vector of known samples with $p<\prod_{i=1}^N n_i$.

To estimate the unknown samples, the authors of \cite{gandy_tensor_2011} propose to solve the following minimization problem, which considers the sums of the nuclear norms of the unfoldings as a convex approximation for the tensor rank:
\begin{align}
    \minimize_{\VX \in \CT} \quad & \sum_{i=1}^{N} \Norm{X_{(i)}}_* + \frac{\lambda}{2} \Norm{\CA(\VX) - \Vb}_2^2,
    \label{eq:nuclear_norm_minimize_unconstrained}
\end{align}
where $\lambda$ is a regularization parameter (we will give more detail on choosing $\lambda$ in \cref{sec:experiments}).

In order to apply the techniques outlined in \cref{sec:dr} to this problem, we first have to compute the proximal maps for the following family of functions:
\begin{align}
    f_i(\VX) &:= \begin{cases}
        \Norm{X_{(i)}}_* &\text{ for } i = 1, \dots, N \\
        \frac{\lambda}{2} \Norm{\CA(\VX) - \Vb}_2^2 &\text{ for } i = N+1.
    \end{cases}
\end{align}

The proximal maps of these functions have already been derived in \cite{gandy_tensor_2011}:
\begin{align}
    &\prox_{\gamma f_i} \VX = \refold(\shrink(X_{(i)}, \gamma)) \qquad \text{for } i=1,\dots,N \label{eq:prox_nuclear_norm}\\
    &(\prox_{\gamma f_{N+1}} \VX)_j = \begin{cases}
        \frac{\gamma}{\lambda\gamma+1}(\lambda \CA^{*} (\Vb) + \frac{1}{\gamma}\VX)_j & \text{if } j \in \Omega \\
        \VX_j & \text{otherwise.}
    \end{cases} \label{eq:prox_Ab}
\end{align}
Here, $\refold \rbdot$ denotes the folding of the matrix into a tensor; i.e. the inverse operation of the unfolding.
The shrinkage operator $\shrink \rbdot$ performs singular value soft-thresholding on a matrix.
First, the singular value decomposition $X = U\Sigma V^{*}$ is computed, where $\Sigma = \diag(\sigma_1(X), \dots, \sigma_r(X))$ is the diagonal matrix containing the singular values of $X$.
Then, $\tilde{\Sigma}= \diag(\max(\sigma_1(X) - \tau, 0), \dots, \max(\sigma_r(X) - \tau, 0))$ is the diagonal matrix containing the singular values shrunk by $\tau$.
The result of the shrinkage operator is then given by $\shrink(X, \tau) := U \tilde{\Sigma} V^{*}$.
The operator $\CA^{*}$ is the adjoint operator of $\CA$, which maps the known entries in the vector $\Vb$ to their corresponding position in the tensor, while setting all unknown entries to zero.

\section{Proposed Algorithms}
\label{sec:algorithm}

When representing a radio map as a matrix, we can expect regular structures in the matrix inherited from buildings, roads, and many other items.
As a result, it is reasonable to assume that rows and columns are in general linearly dependent, making such a matrix rank-deficient.
Therefore, when stacking many matrices representing correlated physical phenomena (e.g. path loss in different heights), we can also expect the resulting tensor to have low rank.
This low-rank property can be exploited for the tensor completion problem by applying the Douglas-Rachford algorithm given by \eqref{eq:dr_splitting} to the problem presented in \cref{sec:rank}.

This approach, however, has severe limitations, since the tensor can be exactly reconstructed only if the number of available samples is large.
This is consistent with \cite{gandy_tensor_2011}, where the authors only provide results for a relatively large number of samples.
For the radio map reconstruction problem at hand, the given number of samples is typically much lower than required by the algorithm presented in \cref{sec:rank}.
In such cases, the algorithm in \cref{sec:rank} produces a (low-rank) solution that is in general no longer consistent with the tensor being estimated.
Severe artifacts might appear because fibers with a very small number of samples, or even with no samples at all, can be replaced by arbitrary copies of other fibers without increasing the rank.
These solutions, however, exhibit a very low spatial coherence, which is not a typical characteristic of radio maps\cite{han2016efficient}, as can be seen in the measurements that are shown later in \cref{fig:madrid_scenario} in \cref{sec:experiments}.

Therefore, in the next sections, we propose two novel algorithms to address these limitations.
First, we add the $\ell_2$ norm of the total variation as regularizer to \eqref{eq:nuclear_norm_minimize_unconstrained} in order to ensure that neighboring tensor elements have similar values.
The second algorithm instead is based on the $\ell_1$ regularization of the total variation in order to reduce the number of discontinuities in the path loss map.

\subsection{L2 Norm Total Variation Regularization}
\label{sec:l2tv}

To take into account that path loss maps have high spatial correlation, we add additional terms to the cost function in \eqref{eq:nuclear_norm_minimize_unconstrained} that penalize abrupt changes in neighboring components of the tensor.
In particular, in this section we use the $\ell_2$ norm of the total variation (Tv) as the penalty term in the first proposed approach.
More precisely, the first approach aims at solving the following problem:
\begin{align}
    \begin{split}
    \minimize_{\VX \in \CT} \quad & \sum_{i=1}^{N} \alpha_i \Tv_2(X_{(i)}) + \sum_{i=1}^{N} \Norm{X_{(i)}}_{*} \\ &+ \frac{\lambda}{2}\Norm{\CA(\VX)-\Vb}_{2}^{2},
    \end{split}
    \label{eq:minimize_tv2}
\end{align}
where $\alpha_i$ and $\lambda$ are regularization parameters, and the $\Tv_2$ operator computes the sum of the squared differences between neighboring elements of each mode-$i$ fiber:
\begin{align}
    \Tv_2 : \IR^{J \times K} \to \IR : X \mapsto \sum_{j=1}^{J} \sum_{k=1}^{K-1} \rbr{x_{j, k+1} - x_{j, k}}^{2},
    \label{eq:tv2}
\end{align}
and where $K = n_i$ and $J = \prod_{\substack{k=1\\k\ne i}}^{N} n_k$.
A scheme for choosing good regularization parameters will be presented in \cref{sec:experiments}.

To apply the algorithm in \cref{sec:dr} to solve \eqref{eq:minimize_tv2}, we first rewrite the problem in \eqref{eq:minimize_tv2} as
\begin{align}
    \minimize_{\VZ \in \CH_0} f(\VZ) + g(\VZ),
    \label{eq:minimize_tv2_rephrased}
\end{align}
where the functions $f$ and $g$ follow the definitions in \eqref{eq:dr_product_f} and \eqref{eq:dr_product_g}, and where the $M = 2N+1$ functions are given by
\begin{align}
    f_i(\VX) &:= \begin{cases}
        \alpha_{i} \Tv_2(X_{(i)}) &\text{ for } i = 1, \dots, N \\
        \Norm{X_{(i-N)}}_* &\text{ for } i = N + 1, \dots, 2N \\
        \frac{\lambda}{2} \Norm{\CA(\VX) - \Vb}_2^2 &\text{ for } i = 2N+1.
    \end{cases}
\end{align}

Using the iterations in \eqref{eq:dr_splitting} to solve \eqref{eq:minimize_tv2_rephrased} requires the proximal maps of $f$ and $g$, which are given in \eqref{eq:prox_f} and \eqref{eq:prox_g}.
This in turn requires the proximal maps of $f_i$ for each $i \in \{1, \dots, 2N+1\}$.
The proximal maps of $f_{N+1}, \dots, f_{2N+1}$ are given in \eqref{eq:prox_nuclear_norm} and \eqref{eq:prox_Ab}, so it remains to derive the proximal maps of $f_{1}, \dots, f_{N}$, which, by definition, are given by
\begin{align}
    \prox_{\gamma f_i} : \VX \mapsto \argmin_{\VY \in \CT} \rbr{ \alpha_{i} \Tv_2(Y_{(i)}) + \frac{1}{2\gamma} \Norm{\VX - \VY}^{2} }.
    \label{eqn:proxD}
\end{align}

The $\Tv_2$ operator is an additively separable function w.r.t. the fibers of the tensor.
Therefore, the proximal map can be computed for each fiber independently \cite{pustelnik2017proximity}:
\begin{align}
    \rbr{\prox_{\gamma f_i} \VX}_{j_1\dots j_{i-1}:j_{i+1}\dots j_N}  = \prox_{\gamma \tilde{f}_i} \rbr{\Vx_{j_1\dots j_{i-1}:j_{i+1}\dots j_N}},
\end{align}
where $\tilde{f}_i(\Vx) = \alpha_i \sum_{k=1}^{K-1} \rbr{x_{k+1} - x_k}^2$ is the penalty term for a single fiber. In turn, the proximal map for $\tilde{f}_i(\Vx)$ is given by
\begin{align}
    \prox_{\gamma \tilde{f}_i} &: \IR^{K} \to \IR^{K}: \Vx \mapsto \argmin_\Vy \rbr{ h_\Vx(\Vy) }, \label{eq:prox_gi}
\end{align}
where
\begin{align}
    h_\Vx(\Vy) & := \alpha_i \sum^{K-1}_{k=1} \rbr{y_{k+1}- y_k}^2 + \frac{1}{2\gamma} \sum_{k=1}^{K} \rbr{x_k - y_k}^2. \label{eq:h_x}
\end{align}

We now proceed to derive the proximal map in \eqref{eq:prox_gi} in closed form. To this end, we first derive the partial derivatives of $h_\Vx(\Vy)$, which are given in \eqref{eqn:prox_der2}, with special cases given in \eqref{eqn:prox_der_1} and \eqref{eqn:prox_der_N}:
\begin{align}
    \frac{\partial h_\Vx(\Vy)}{\partial y_j} &= \rbr{4\alpha_i + \frac{1}{\gamma}} y_j - 2\alpha_i y_{j-1} - 2\alpha_i y_{j+1} - \frac{1}{\gamma} x_j
    \label{eqn:prox_der2}, \\
    \frac{\partial h_\Vx(\Vy)}{\partial y_1} &= \rbr{4 \alpha_i + \frac{1}{\gamma}} y_1 - 2\alpha_i y_2 - \frac{1}{\gamma }x_1
    \label{eqn:prox_der_1}, \\
    \frac{\partial h_\Vx(\Vy)}{\partial y_K} &= \rbr{4 \alpha_i + \frac{1}{\gamma}} y_K - 2\alpha_i y_{K-1} - \frac{1}{\gamma} x_K
    \label{eqn:prox_der_N}.
\end{align}

Since \eqref{eq:h_x} is convex and the problem is unconstrained, we can find a global minimizer by setting all partial derivates to zero to obtain the following system of linear equations:
\begin{equation}
    A \Vy = \frac{1}{\gamma}\Vx \label{eqn:lin_sys},
\end{equation}
where
\begin{equation*}
    A = \begin{bmatrix}
    (4\alpha_i+ \frac{1}{\gamma}) & -2 \alpha_i & 0 & \dots & 0 \\
    -2 \alpha_i & (4\alpha_i+ \frac{1}{\gamma}) & -2 \alpha_i & \dots & 0 \\
    0 & -2 \alpha_i & (4\alpha_i+ \frac{1}{\gamma}) & \dots & 0 \\
    \vdots & \vdots & \vdots & \ddots & \vdots \\
    0 & 0 & 0 & \dots & (4\alpha_i+ \frac{1}{\gamma})
    \end{bmatrix}
\end{equation*}

The above system of linear equations can be solved efficiently because $A$ is a tridiagonal Toeplitz matrix, which can be efficiently inverted\cite{dow2008explicit}, so the following vector is easy to obtain in practice:
\begin{align}
    \Vy &= \frac{1}{\gamma}A\inv \Vx.
    \label{eqn:lin_sys_inv}
\end{align}

The result of applying this equation to all fibers of the tensor can be written more compactly as the $i$-mode multiplication of the tensor with $A\inv$:
\begin{align}
    \prox_{\gamma f_i} : \VX \mapsto \VX \times_{i} \frac{1}{\gamma} A\inv. \label{eq:prox_tv2}
\end{align}

We have now derived the proximal map of $f_i$ for each $i \in \{1, \dots, 2N+1\}$ in \eqref{eq:prox_nuclear_norm}, \eqref{eq:prox_Ab}, and \eqref{eq:prox_tv2}, which, together with \eqref{eq:prox_f} and \eqref{eq:prox_g}, enables us to apply the Douglas-Rachford splitting method to solve \eqref{eq:minimize_tv2_rephrased}.

As we show later by simulations, since this algorithm uses additional information, it is able to reconstruct the tensor accurately with fewer samples compared to the algorithm presented in \cref{sec:rank}.
One potential limitation of the algorithm presented in this section is that transitions provoked by, for example, building edges, might be smoothed heavily due to the quadratic penalty of the $\ell_2$ norm.

\begin{remark}
In numerical experiments, we have noticed that the magnitude of the edges of the tensor decreases in each iteration. To alleviate this effect, we propose, as a heuristic, to scale each row of the matrix $A\inv$ to have a sum of 1.
\label{rem:heuristic}
\end{remark}

\subsection{L1 Norm Total Variation Regularization}
\label{sec:l1tv}

To overcome the potential limitation of the algorithms introduced in the previous subsection, we now present an algorithm that uses an $\ell_1$ regularizer in place of the $\ell_2$ regularizer.
Intuitively, by using the $\ell_1$ norm, we minimize the number of sharp transitions between neighboring elements of the tensor unfoldings.
Formally, the $\ell_1$ norm total variation regularization problem is given as follows:
\begin{equation}
    \begin{split}
    \minimize_{\VX \in \CT} \quad & \sum_{i=1}^{N} \alpha_i \Tv_1(X_{(i)}) + \sum_{i=1}^{N} \Norm{X_{(i)}}_{*} \\ &+ \frac{\lambda}{2}\Norm{\CA(\VX)-\Vb}_{2}^{2},
    \end{split}
    \label{eq:minimize_tv1}
\end{equation}
where $\alpha_i$ and $\lambda$ are regularization parameters, and
\begin{equation}
    \Tv_1 (X) := \sum_{j=1}^{J} \sum_{k=1}^{K-1} \Abs{x_{j, k+1} - x_{j, k}}.
\end{equation}

To solve \eqref{eq:minimize_tv1}, we can again use the algorithm presented in \cref{sec:dr} with the following definition:
\begin{align}
    f_i(\VX) &:= \begin{cases}
        \alpha_{i} \Tv_1(X_{(i)}) &\text{ for } i = 1, \dots, N \\
        \Norm{X_{(i-N)}}_* &\text{ for } i = N + 1, \dots, 2N \\
        \frac{\lambda}{2} \Norm{\CA(\VX) - \Vb}_2^2 &\text{ for } i = 2N+1.
    \end{cases}
\end{align}

The challenge in solving \eqref{eq:minimize_tv1} with the Douglas-Rachford splitting method is the computation of the proximal maps for $f_{1}, \dots, f_{N}$.
The method presented in \cref{sec:l2tv} is not applicable here since the absolute value function is non-differentiable.
Typically, in these cases the proximal maps are computed with iterative methods such as those described in \cite{barbero2014modular}, which is the approach we use in the simulations.
However, some authors have also proposed to use neural networks as fast approximation of proximal maps\cite{sprechmann2015learning}.

\section{Numerical Experiments and Conclusion}
\label{sec:experiments}

\begin{figure}[tb]
    \centering
    \includegraphics[width=0.75\linewidth]{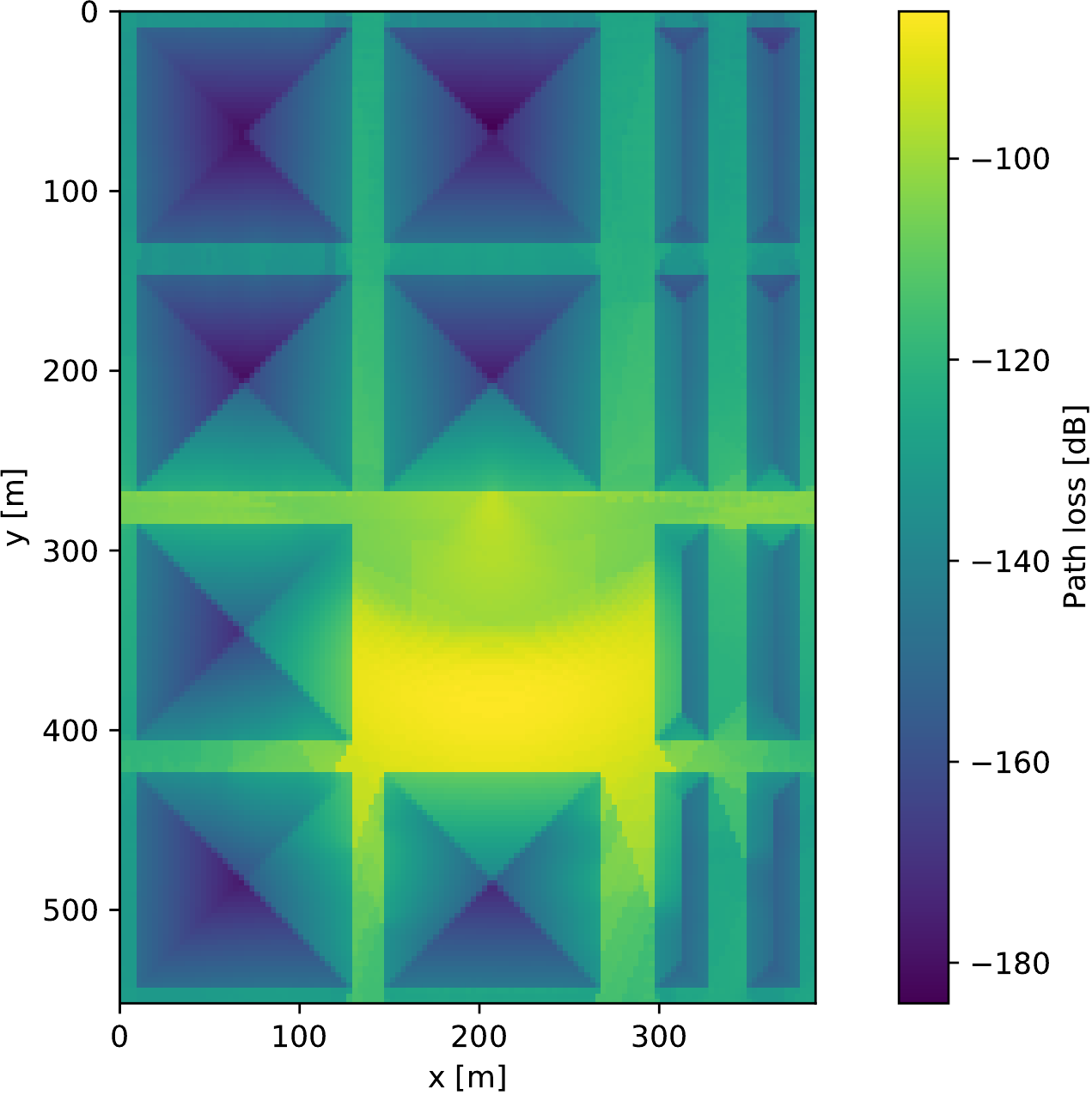}
    \caption{Path loss for Madrid scenario at height of \SI{1.5}{m}.}
    \label{fig:madrid_scenario}
\end{figure}

To evaluate the performance of the algorithms, we used synthetic path loss measurements obtained by ray tracing simulations during the METIS2020 project\cite{metis_d61}.
The chosen scenario (Madrid scenario) represents an urban environment with buildings of different heights, parks and streets, all arranged on a regular grid.
The path loss measurements for a height of \SI{1.5}{m} are shown in \cref{fig:madrid_scenario}, but measurements for \SI{26}{m} and \SI{43.5}{m} were also available.

The available measurements\footnote{available at https://metis2020.com/wp-content/uploads/simulations/Ray-Tracing-files-for-TC2-Part-I.zip} contain three repetitions of the environment model in $x$ and $y$ direction to avoid border effects.
For the simulations, however, the repetitions were removed, and only the central block was used.
The resolution of the measurements was \SI{3}{m}, and the area had a size of $\SI{387}{m} \times \SI{552}{m}$, resulting in a tensor of dimensions $129 \times 184 \times 3$.

For the simulations, we randomly selected a given percentage of tensor elements, and we used these as input to the algorithms.
As the performance metric, we used the normalized mean squared error (NMSE), which is defined as
\begin{align}
    \text{NMSE}_\text{dB} = 10 \log_{10}\rbr{\frac{\sum\limits_{j \in \Omega_\mathrm{m}} \rbr{\hat{x}_j - x_j}^2}{\sum\limits_{j \in \Omega_\mathrm{m}} x_j^2}}, \label{eq:nmse}
\end{align}
where $\hat{\VX}$ and $\VX$ are reconstructed and correct tensors given in \si{dB}, respectively, and $\Omega_\mathrm{m} \subseteq \cbr{1, \dots, n_1} \times \dots \times \cbr{1, \dots, n_N}$ is the set of indizes that were not used as input to the algorithms.

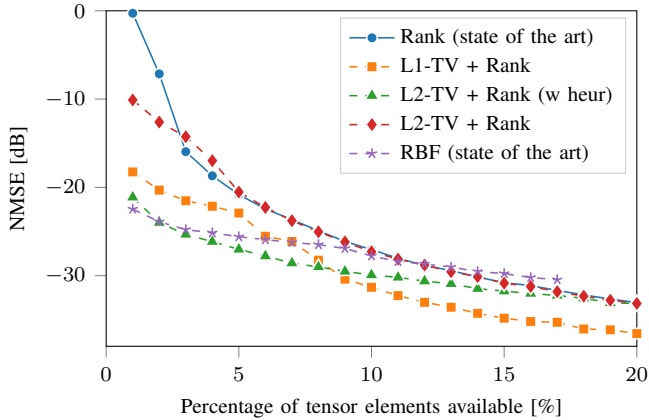
\begin{figure}[tb]
    \centering
    \newlength\figureheight
    \newlength\figurewidth
    \setlength\figureheight{0.7\linewidth}
    \setlength\figurewidth{\linewidth}
    \tikzset{font={\footnotesize}}
    \input{results-madrid.tikz}
    \vspace{-0.5cm}
    \caption{Reconstruction error for different algorithms.}
    \label{fig:results_madrid}
\end{figure}

The regularization parameters $(\alpha_i)_{i \in \cbr{1, \dots, N}}$ were obtained by using cross-validation with the holdout method.
More specifically, \SI{25}{\%} of the given tensor elements were removed from the measurements at random and then used as test set.
We then selected a set of reasonable regularization parameter values and applied cross-validation to determine the best parameters out of all possible combinations.
For given regularization parameters $\alpha_i$ we applied the continuation technique described in \cite{gandy_tensor_2011} to $\lambda$ to achieve a sufficiently small value for $\Norm{\CA(\VX) - \Vb}_2$.
To this end, we ran the algorithm for a small value for $\lambda$ until no significant progress was achieved, then increased $\lambda$ and ran the algorithm again.
We repeated this process until either a further increase of $\lambda$ had no influence on the reconstruction result or a given number of iterations was achieved.

For comparison, we implemented the radial basis function (RBF) algorithm, as described in \cite[Sect. 5.1]{bishop1995neural}. We achieved the best results with the multiquadric function $\phi(r) = \sqrt{1 + \rbr{\frac{r}{\epsilon}}^2}$, where the hyperparameter $\epsilon$ was chosen by line search, using the same cross-validation procedure as described for the other algorithms.

As can be seen in \cref{fig:results_madrid}, the algorithm presented in \cref{sec:l1tv} (denoted as \emph{L1-TV + Rank}) yields a constant improvement over the existing algorithm presented in \cref{sec:rank} (denoted as \emph{Rank}).
On the other hand, the algorithm presented in \cref{sec:l2tv} with the heuristic mentioned in \cref{rem:heuristic} (denoted as \emph{L2-TV + Rank (w heur)}), although not particularly good for a large number of samples, provides a significant improvement for a smaller number of samples.
Only when less than \SI{2}{\percent} of all tensor elements are given for reconstruction is it slightly outperformed by the RBF algorithm.

The fact that the proposed algorithms greatly outperform state-of-the-art techniques indicates that the proposed algorithms are successfully exploiting the assumed structure of radio maps.
The comparison to the RBF algorithm shows that the assumption of smoothness is not sufficient and exploiting the low-rank property is necessary to achieve good performance in a wide range of conditions.
The proposed algorithms form a framework which is very flexible and can be applied to a wide range of problems in different areas.

\bibliographystyle{IEEEtran}
\bibliography{IEEEabrv,radiomaps}

\end{document}

%% file: results-madrid.tikz
\begin{tikzpicture}

\definecolor{color0}{rgb}{0.12156862745098,0.466666666666667,0.705882352941177}
\definecolor{color1}{rgb}{1,0.498039215686275,0.0549019607843137}
\definecolor{color2}{rgb}{0.172549019607843,0.627450980392157,0.172549019607843}
\definecolor{color3}{rgb}{0.83921568627451,0.152941176470588,0.156862745098039}
\definecolor{color4}{rgb}{0.580392156862745,0.403921568627451,0.741176470588235}

\begin{axis}[
height=\figureheight,
legend cell align={left},
legend entries={{Rank (state of the art)},{L1-TV + Rank},{L2-TV + Rank (w heur)},{L2-TV + Rank},{RBF (state of the art)}},
legend style={draw=white!80.0!black},
tick align=outside,
tick pos=left,
width=\figurewidth,
x grid style={white!69.01960784313725!black},
xlabel={Percentage of tensor elements available [\%]},
xmin=0, xmax=20,
y grid style={white!69.01960784313725!black},
ylabel={NMSE [dB]},
ymin=-38, ymax=0
]

\addplot [semithick, color0, mark=*, mark size=2, mark options={solid,draw=white}]
table [row sep=\\]{%
1	-0.27791416521067 \\
2	-7.14480843772342 \\
3	-15.956950421304 \\
4	-18.6882888911373 \\
5	-20.8216051346727 \\
6	-22.4383896219879 \\
7	-23.6554833651957 \\
8	-24.9121433596875 \\
9	-26.1287995590375 \\
10	-27.0449084448455 \\
11	-28.0384409837157 \\
12	-28.845343444154 \\
13	-29.4268532486362 \\
14	-30.1504200506176 \\
15	-30.8029476068079 \\
16	-31.1229089979704 \\
17	-31.6739184220415 \\
18	-32.1836043496255 \\
19	-32.6533034740927 \\
20	-33.048992529831 \\
};
\addplot [semithick, color1, dashed, mark=square*, mark size=2, mark options={solid,draw=white}]
table [row sep=\\]{%
1	-18.2467012686982 \\
2	-20.3137700236786 \\
3	-21.5128292466481 \\
4	-22.1379543973401 \\
5	-22.9094487933738 \\
6	-25.5391540495825 \\
7	-26.1288554977225 \\
8	-28.2542021155079 \\
9	-30.4004471656369 \\
10	-31.3185281325448 \\
11	-32.2608712684598 \\
12	-33.0196227748747 \\
13	-33.5772596657437 \\
14	-34.2665179239188 \\
15	-34.8199168497624 \\
16	-35.1913718648454 \\
17	-35.2837798993128 \\
18	-36.0368843758319 \\
19	-36.1350217281837 \\
20	-36.5678786615508 \\
};
\addplot [semithick, color2, dashed, mark=triangle*, mark size=3, mark options={solid,draw=white}]
table [row sep=\\]{%
1	-21.1245295831946 \\
2	-23.9884818278635 \\
3	-25.3319238923736 \\
4	-26.1486701015737 \\
5	-27.0227459361372 \\
6	-27.7812725319179 \\
7	-28.5753485953252 \\
8	-29.0129231014929 \\
9	-29.5228729845351 \\
10	-29.923974260914 \\
11	-30.1951627098689 \\
12	-30.6216103765658 \\
13	-30.9183979107652 \\
14	-31.4741435291525 \\
15	-31.7571970409311 \\
16	-32.024058467465 \\
17	-32.2463267728153 \\
18	-32.5599236778567 \\
19	-33.0860094332124 \\
20	-33.1477233946208 \\
};
\addplot [semithick, color3, dashed, mark=diamond*, mark size=3, mark options={solid,draw=white}]
table [row sep=\\]{%
1	-10.1011852680647 \\
2	-12.6046950790946 \\
3	-14.2661110964754 \\
4	-16.9742878360023 \\
5	-20.5278943126415 \\
6	-22.2754407356523 \\
7	-23.7740446470514 \\
8	-25.0342415911949 \\
9	-26.1816402822879 \\
10	-27.2952624052892 \\
11	-28.1247203512746 \\
12	-28.8030987261202 \\
13	-29.5555459460425 \\
14	-30.1069847275437 \\
15	-30.8560769984788 \\
16	-31.2201735746927 \\
17	-31.8197372920881 \\
18	-32.3244774319506 \\
19	-32.7822290122103 \\
20	-33.1510160194483 \\
};
\addplot [semithick, color4, dashed, mark=star, mark size=2, mark options={solid,draw=color4}]
table [row sep=\\]{%
1	-22.4580197668039 \\
2	-23.8925542726553 \\
3	-24.7889474046822 \\
4	-25.1500879040937 \\
5	-25.5967944241462 \\
6	-25.9183013481434 \\
7	-26.2387122566897 \\
8	-26.4868851650512 \\
9	-26.921548585555 \\
10	-27.7708427538918 \\
11	-28.3932514042878 \\
12	-28.6873481812273 \\
13	-29.0308584678588 \\
14	-29.5334145964976 \\
15	-29.7605475949918 \\
16	-30.2112303911362 \\
17	-30.4654561985288 \\
};
\end{axis}

\end{tikzpicture}